# Minimum-Information LQG Control
# Part I: Memoryless Controllers

Roy Fox[†] and Naftali Tishby[†]

*Abstract*— With the increased demand for power efficiency in feedback-control systems, communication is becoming a limiting factor, raising the need to trade off the external cost that they incur with the capacity of the controller's communication channels. With a proper design of the channels, this translates into a sequential rate-distortion problem, where we minimize the rate of information required for the controller's operation under a constraint on its external cost. Memoryless controllers are of particular interest both for the simplicity and frugality of their implementation and as a basis for studying more complex controllers. In this paper we present the optimality principle for memoryless linear controllers that utilize minimal information rates to achieve a guaranteed external-cost level. We also study the interesting and useful phenomenology of the optimal controller, such as the principled reduction of its order.

## I. INTRODUCTION

The modern technology industry is deploying artificial sensing-acting agents everywhere [1]. From smart-home devices to manufacturing robots to outdoor vehicles and from nanoscale machines to space rockets, these agents sense their environment and act on it in a perception-action cycle [2].

When these agents are centrally controlled or when the sensors and the actuators are distributed, this control process relies on the ability to communicate the observations to the controller and the intended actions to the actuators. Autonomous agents likewise require sufficient capacity for the internal communication between their sensor and actuator components. As devices become smaller and more ubiquitous, power efficiency and physical restrictions dictate that communication become a limiting factor in the agent's operation.

Classic optimal control theory [3] is unconcerned with the costs and the limitations of communicating the information needed for the controller's operation. In the past two decades, however, a large body of research has been dedicated to this issue ([4]–[7] and references therein).

The perception-action cycle between a controller and its environment (Figure 1) consists of multiple channels and the capacity of any of them can be limited. Accordingly, various information rates can be considered. Our guiding principle in this work is to measure the information complexity of the controller's internal representation by asking *"How much information does the controller have on the past?"*. The past is informative of the future [8] and some information in past observations is useful in controlling the future. We therefore seek a trade-off between the external cost incurred by the system and the internal cost of the communication resources spent by the controller in reducing that external cost. This trade-off is often formulated as an optimization problem, where one cost is constrained and the other minimized.

When the controller has no internal memory, it can only attend to its most recent input observation, perhaps selectively. The degree of this attention, measured by the amount of Shannon information about the input observation that is utilized in the output control, is a lower bound on the required capacity of the communication channel between the controller's sensor and its actuator (see Figure 3).

Our motivation in considering memoryless controllers is twofold. First, there are applications in which having any significant memory capacity within the controller is impractical. When the system is complex and the controller's hardware and resources are limited, they may be inadequate for maintaining any significant representation of the environment. In this case, a memoryless controller is the more cost-effective solution and sometimes the only feasible one. Memoryless controllers have been studied before, particularly in the contexts of delay [9]–[11] and discrete state-spaces [12]–[14].

Second, we show in Part II of this work [15] how to formulate the problem of optimizing a bounded retentive (memory-utilizing) controller as an equivalent problem of optimizing a bounded memoryless controller. This reduction enables us to reuse the solution derived in this paper in solving the bounded retentive control problem.

Much of the related existing research has been concerned with the issue of stabilizability of an unstable plant over communication channels that are limited in some way: quantization [16]–[20], noise [21]–[23], delay [24] and fading [25]. Our current work reduces in the stabilizable case to known results, and this analysis will be included in an upcoming paper.

Other early publications proposed heuristic approximate solutions to the problem of optimal control with finite precision [26], [27]. More recently, the problem of optimal control over limited-capacity channels has been studied, with various information patterns in the sensor-side encoder and the actuator-side decoder: unlimited encoder and decoder memory with full feedback [28]–[31], unlimited encoder memory and memoryless decoder [32], and unlimited decoder memory with some feedback to the encoder [33].

A special case of our current work was studied in [34]. Their setting is fully observable and scalar, whereas we treat the much more general setting of partially observable vector spaces. Our main result reduces in this simple case to one of their solutions, implying that their other proposed solution is never optimal.

---

[†]School of Computer Science and Engineering, The Hebrew University, {royf,tishby}@cs.huji.ac.il
*This work was supported by the DARPA MSEE Program, the Gatsby Charitable Foundation, the Israel Science Foundation and the Intel ICRI-CI Institute

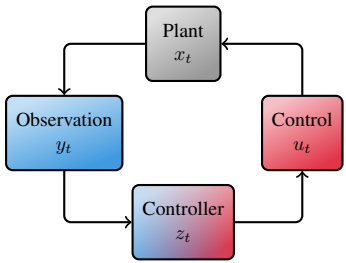

Fig. 1. Block diagram of a closed-loop control system

In this paper we make three contributions. First, we present a method for designing memoryless linear controllers that utilize minimal information rates to achieve a guaranteed external cost level. To our knowledge, this is the first treatment of information considerations in continuous-space control problems where neither the controller's sensor nor its actuator have unbounded memory capacity.

Second, we derive a solution that has a particularly explicit form, allowing direct numerical computation. Unlike classic controllers, which are designed by separable forward and backward Riccati equations [35], our forward and backward recursions are coupled. Yet each forward and backward step is given in closed form up to eigenvalue decomposition (EVD) operations. This is in contrast to the semidefinite programs (SDP) in [29], [31], [36], which require external solvers.

Third, we study the interesting and useful phenomenology of the optimal controller. It manifests a water-filling effect [37], which is a principled criterion for the selection of the active controller modes and their magnitudes. By trading off external cost to reduce the controller's communication resources we also reduce its order in a principled way.

In Section II we define the LQG task that the controller should solve. In Section III we present the memoryless control model and the information considerations involved. In Section IV we find the conditions satisfied by the optimal linear solution and discuss its intriguing phenomenology. More discussion and an illustrative example can be found in Part II of this work [15].

## II. CONTROL TASK

We consider the closed-loop control problem depicted schematically in Figure 1, where an agent (controller) is interacting with its environment (plant). When the plant is in state $x_t \in \mathbb{R}^n$, it emits an observation $y_t \in \mathbb{R}^k$, takes in a control input $u_t \in \mathbb{R}^\ell$ and undergoes a stochastic state transition. The goal of the controller is to reduce the long-term average expectation of some cost rate $\mathcal{J}_t(x_t, u_t)$.

A controller $\pi$ defines the possibly stochastic mapping from the observable history $y^t = \{y_\tau\}_{\tau \leq t}$ into the control $u_t$. The plant and the controller, under some initial conditions, jointly induce a stochastic process over the infinite sequence of variables $\{x_t, y_t, u_t\}$.

Our focus in this work is on discrete-time systems with linear dynamics, Gaussian noise and quadratic cost rate (LQG). For simplicity, all elements are taken to be homogeneous, i.e. centered at the origin, and time-invariant. We note that all our results hold without these assumptions, with the appropriate adjustments, as usual in LQG problems [3].

*Definition 1:* A linear-Gaussian time-invariant (LTI) plant $\langle A, B, C, \Sigma_\xi, \Sigma_\epsilon \rangle$ has state dynamics

$$x_{t+1} = Ax_t + Bu_t + \xi_t; \qquad \xi_t \sim \mathcal{N}(0, \Sigma_\xi),$$

where $A \in \mathbb{R}^{n \times n}$, $B \in \mathbb{R}^{n \times \ell}$, $\Sigma_\xi \in \mathbb{S}^n_+$ is in the positive-semidefinite cone and $\xi_t$ is independent of $(x^t, y^t, u^t) = \{x_\tau, y_\tau, u_\tau\}_{\tau \leq t}$. The observation dynamics are

$$y_t = Cx_t + \epsilon_t; \qquad \epsilon_t \sim \mathcal{N}(0, \Sigma_\epsilon),$$

where $C \in \mathbb{R}^{k \times n}$, $\Sigma_\epsilon \in \mathbb{S}^k_+$ and $\epsilon_t$ is independent of $(y^{t-1}, u^{t-1}, x^t)$.

*Definition 2:* A linear-quadratic-Gaussian (LQG) task $\langle A, B, C, \Sigma_\xi, \Sigma_\epsilon, Q, R \rangle$ involves a LTI plant and the cost rate

$$\mathcal{J}_t = \tfrac{1}{2}(x_t^\intercal Q x_t + u_t^\intercal R u_t),$$

where $Q \in \mathbb{S}^n_+$ and $R \in \mathbb{S}^\ell_+$. The task is to achieve a low long-term average expected cost rate, with respect to the distribution induced by the plant and the controller $\pi$

$$\mathcal{J}_\pi = \limsup_{T \to \infty} \frac{1}{T} \sum_{t=1}^T \mathbb{E}_\pi[\mathcal{J}_t]. \qquad (1)$$

We are particularly interested in controllers which are time-invariant, i.e. have $\pi(u_t|y^t)$ independent of $t$, and which induce a stationary process, independent of any initial conditions. In a stationary process, the marginal joint distribution of $(x_t, y_t, u_t)$ is time-invariant and we can replace the long-term average expected cost rate (1) with the expected cost rate in the stationary marginal distribution.

We denote by $\Sigma_x \in \mathbb{S}^n_+$ and $\Sigma_y \in \mathbb{S}^k_+$, respectively, the stationary covariances of the state and of the observation, assuming they exist and are finite. They are related through

$$\Sigma_y = C \Sigma_x C^\intercal + \Sigma_\epsilon.$$

If $x_t$ and $y_t$ are jointly Gaussian with mean 0, they satisfy the reverse relation

$$x_t = K y_t + \kappa_t; \qquad \kappa_t \sim \mathcal{N}(0, \Sigma_\kappa),$$

where the residual state noise $\kappa_t$ is independent of $y_t$ (but not of the past of the process), and

$$K = \Sigma_x C^\intercal \Sigma_y^\dagger$$
$$\Sigma_\kappa = \Sigma_x - \Sigma_x C^\intercal \Sigma_y^\dagger C \Sigma_x,$$

with $\cdot^\dagger$ the Moore-Penrose pseudoinverse. If the entire process has mean 0, the stationary expected cost rate (1) is given by

$$\mathcal{J}_\pi = \tfrac{1}{2}(\mathrm{tr}(Q \Sigma_x) + \mathrm{tr}(R \Sigma_u)), \qquad (2)$$

where $\Sigma_u \in \mathbb{S}^\ell_+$ is the stationary control covariance.

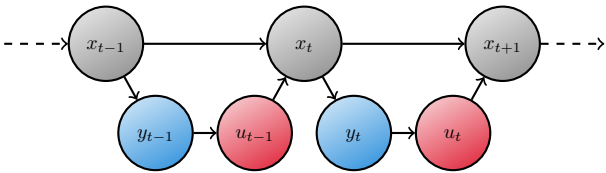

Fig. 2. Bayesian network of memoryless control

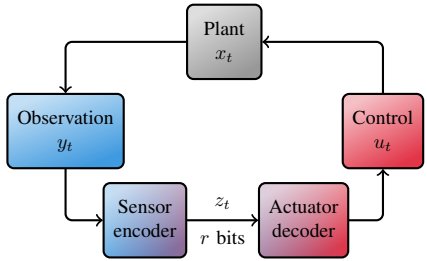

Fig. 3. The communication channel from the sensor to the actuator

## III. BOUNDED MEMORYLESS CONTROLLERS

### A. Control model

In this section we introduce memoryless controllers with bounded communication resources. A memoryless controller is simply a possibly stochastic mapping from its input observation $y_t$ into its output control $u_t$ without any memory of past observations.

*Definition 3:* A controller is memoryless if the control depends only on the most recent observation; that is, $u_t$ is independent of $(y^{t-1}, u^{t-1}, x^t)$ given $y_t$.

A system including a memoryless controller satisfies the Bayesian network in Figure 2.

Optimization over the space of all measurable control laws is hard to analyze and the optimal controller can be hard to implement. It is therefore practical to require the control law to have a certain form, most commonly the linear-Gaussian time-invariant (LTI) form. LTI controllers induce, jointly with a LTI plant, a Gaussian stochastic process. When the process is stable, it has a unique stationary distribution that is independent of any initial conditions. Linear controllers with limited memory are known not to be optimal for all control problems [38], [39]. The conditions under which there exists an optimal memoryless controller which is LTI, so that no performance is lost by focusing our attention on such controllers, are beyond the scope of this paper.

*Definition 4:* A memoryless linear-Gaussian time-invariant (LTI) controller has control law of the form

$$u_t = Hy_t + \eta_t; \qquad \eta_t \sim \mathcal{N}(0, \Sigma_\eta), \qquad (3)$$

where $H \in \mathbb{R}^{\ell \times k}$, $\Sigma_\eta \in \mathbb{S}_+^\ell$ and $\eta_t$ is independent of $y_t$.

### B. Information considerations

Our controller is bounded and operates under limitations on its capacity to process the observation and produce the control. To measure this internal complexity of the controller, we consider a memoryless communication channel from the sensor to the actuator with limited capacity (Figure 3).

For example, we can consider a noiseless binary channel and measure the controller complexity by the number $r$ of bits per time step that it transmits from its sensor to its actuator. This requires the controller's sensor to perform lossy source coding of the observation $y_t$ by compressing it into a binary string representation $z_t \in \{0,1\}^r$. This representation is transmitted losslessly and reconstructed by the controller's actuator as a control $u_t$. Since the controller is memoryless, both the encoder and the decoder are memoryless.

In this sense, the dynamical control problem can be thought of as a sequential rate-distortion (SRD) problem [28], [36]. Unlike the standard one-shot rate-distortion (RD) problem [37], [40], in a SRD problem the output distribution affects the future of the process. This often creates a coupling between the forward inference process that determines the marginal distributions and the backward control process that determines the cost-to-go, i.e. the distortion. We note that without control [36] the decoder only affects the controller part of the future trajectory; however, this distinction is of minor consequence for the SRD aspect of the problem [41].

Following rate-distortion theory, we find that the bit rate $r$ required for this process is linked to the Shannon mutual information between the observation and the control, defined by

$$\mathbb{I}[y_t; u_t] = \mathbb{E}\left[\log \frac{f(y_t, u_t)}{f(y_t)f(u_t)}\right],$$

where $f$ denotes the various probability density functions, as indicated by their arguments. The bit rate is bounded from below by the information rate due to the data-processing inequality [37]

$$\mathbb{I}[y_t; u_t] \leq \mathbb{I}[y_t; z_t] \leq \mathbb{H}[z_t] \leq r \log 2,$$

where

$$\mathbb{H}[z_t] = -\mathbb{E}[\log \Pr(z_t)]$$

is the discrete Shannon entropy of $z_t$.

In classic information theory, this bound is made asymptotically tight by jointly encoding a long block of observations and jointly decoding a long block of controls. In our setting, this is impossible due to the causal nature of the plant-controller interaction. Thus, unfortunately, the bound is generally not tight for discrete channels. We can nevertheless expect it to be a good approximation, if we draw intuition from the stabilizability problem, where the informational lower bound is approximated by a known upper bound [42].

In applications, it is often possible to make design choices regarding the channel itself. If we can design the channel to be perfectly matched to the optimal LTI control law, no block coding will be needed [43]. When the controller is LTI, it is more practical to take the channel in Figure 3 to be itself linear-Gaussian instead of binary. There exists an additive Gaussian noise channel with a signal power cost that is perfectly matched to our optimal controller in Theorem 1. With such a channel, the information rate is optimally equal to the channel capacity and a constraint on the information rate $\mathbb{I}[y_t; u_t]$ is equivalent to a constraint on the expected power available for transmission on the channel. We develop these results in the Supplementary Material[1] (SM), Appendix I.

We are thus interested in a LTI controller $\pi$ that minimizes the long-term average

$$\mathcal{I}_\pi = \limsup_{T \to \infty} \frac{1}{T} \sum_{t=1}^T \mathcal{I}_t \quad (4)$$

of the controller's internal information rate $\mathcal{I}_t = \mathbb{I}[y_t; u_t]$, under the constraint that it achieves some guarantee level $c$ of expected cost rate.

*Problem 1:* Given a LQG task, the bounded memoryless LTI controller optimization problem is

$$\min_\pi \quad \mathcal{I}_\pi$$
$$\text{s.t.} \quad \mathcal{J}_\pi \leq c,$$

with $\mathcal{I}_\pi$ as in (4), where $\mathcal{I}_t = \mathbb{I}[y_t; u_t]$, and with $u_t$ as in (3).

## IV. MAIN RESULT

### A. Optimality conditions

In this section we derive the optimality conditions for a bounded memoryless LTI controller. These conditions are summarized in Theorem 1 below.

Analysis of Problem 1 starts with considering the minimum mean square error (MMSE) estimators

$$\hat{x}_{y_t} = \mathbb{E}[x_t|y_t] = K y_t$$
$$\hat{x}_{u_t} = \mathbb{E}[x_t|u_t] = \Sigma_{x;u} \Sigma_u^\dagger u_t,$$

respectively for the state given the observation and the control. Here $\Sigma_{x;u} = \mathbb{E}[x_t u_t^\intercal]$ is the covariance matrix between $x_t$ and $u_t$. This implies that $\hat{x}_{y_t}$ and $\hat{x}_{u_t}$ are also 0-mean and jointly Gaussian with the other variables. At this point, it is useful to state a few properties of MMSE estimators of Gaussian variables.

*Lemma 1:* Let $x$ and $\hat{x}$ be 0-mean jointly Gaussian random variables. The following properties are equivalent:
1) There exists a random variable $u$, jointly Gaussian with $x$, such that $\hat{x}(u) = \arg\min_{\hat{x}} \mathbb{E}[\|\hat{x} - x\|^2|u] = \mathbb{E}[x|u]$.
2) $\Sigma_{\hat{x};x} = \Sigma_{\hat{x}}$.
3) $\Sigma_{x|\hat{x}} = \Sigma_x - \Sigma_{\hat{x}}$, where $\Sigma_{x|\hat{x}}$ is the conditional covariance matrix of $x$ given $\hat{x}$, implying $\Sigma_x \succeq \Sigma_{\hat{x}}$.
4) $\hat{x} = \mathbb{E}[x|\hat{x}]$.

Such $\hat{x}$ is called a minimum mean square error (MMSE) estimator (of $u$) for $x$.

*Proof:* See SM, Appendix II. $\square$

Since the conditional covariance $\Sigma_{x|u}$ of $x_t$ given $u_t$ is deterministic, i.e. is not a random variable, the conditional expectation of $x_t$ given $u_t$, i.e. $\hat{x}_{u_t}$, is a sufficient statistic of $u_t$ for $x_t$, satisfying the Markov chain $x_t - \hat{x}_{u_t} - u_t$. This suggests that the stochastic control process satisfies the Bayesian network in Figure 4, where the control is based on $\hat{x}_{u_t}$ instead of directly on $y_t$.

[1]Available at https://arxiv.org/abs/1946

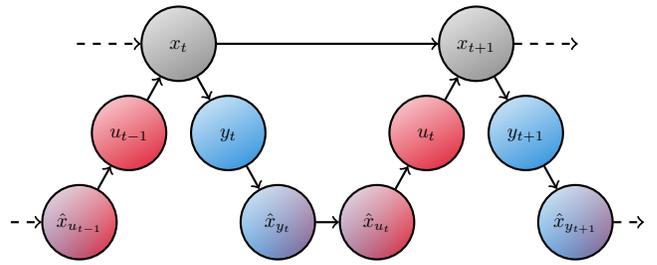

Fig. 4. Bayesian network of memoryless estimator-based control

*Lemma 2:* The bounded memoryless LTI controller optimization problem (Problem 1) is solved by a control law of the form

$$\hat{x}_{y_t} = K y_t \quad (5a)$$
$$\hat{x}_{u_t} = W \hat{x}_{y_t} + \omega_t; \quad \omega_t \sim \mathcal{N}(0, \Sigma_\omega) \quad (5b)$$
$$u_t = L \hat{x}_{u_t}, \quad (5c)$$

where $W \in \mathbb{R}^{n \times n}$, $\Sigma_\omega \in \mathbb{R}^{n \times n}$, $L \in \mathbb{R}^{\ell \times n}$, $\omega_t$ is independent of $y_t$, $\hat{x}_{u_t}$ is a MMSE estimator for $\hat{x}_{y_t}$ and

$$\mathbb{I}[y_t; u_t] = \mathbb{I}[\hat{x}_{y_t}; \hat{x}_{u_t}]. \quad (6)$$

*Proof:* See SM, Appendix III. $\square$

Lemma 2 allows us to derive optimality conditions for Problem 1. The stationary state covariance satisfies

$$\Sigma_x = \begin{bmatrix} A & B \end{bmatrix} \begin{bmatrix} \Sigma_x & \Sigma_{x;u} \\ \Sigma_{u;x} & \Sigma_u \end{bmatrix} \begin{bmatrix} A & B \end{bmatrix}^\intercal + \Sigma_\xi \quad (7)$$
$$= (A + BL) \Sigma_{\hat{x}_u} (A + BL)^\intercal + A \Sigma_{x|\hat{x}_u} A^\intercal + \Sigma_\xi.$$

The mutual information between jointly Gaussian variables [37] is given by

$$\mathbb{I}[\hat{x}_{y_t}; \hat{x}_{u_t}] = \tfrac{1}{2}(\log |\Sigma_{\hat{x}_y}|_\dagger - \log |\Sigma_{\hat{x}_y|\hat{x}_u}|_\dagger), \quad (8)$$

where $|\cdot|_\dagger$ is the pseudodeterminant, i.e. the product of the positive eigenvalues. This holds if $\Sigma_{\hat{x}_y}$ and $\Sigma_{\hat{x}_y|\hat{x}_u}$ have the same range and thus the same number of positive eigenvalues; otherwise, the mutual information between $y_t$ and $u_t$ is infinite.

With the target (8) and the constraints (7) and $\mathcal{J}_\pi \leq c$, where $\mathcal{J}_\pi$ is given by (2), the Lagrangian of Problem 1 can be written as

$$\mathcal{F}_{\Sigma_x, \Sigma_{\hat{x}_u}, L, S; \beta} = \tfrac{1}{2}(\beta^{-1}(\log |\Sigma_{\hat{x}_y}|_\dagger - \log |\Sigma_{\hat{x}_y|\hat{x}_u}|_\dagger) \quad (9)$$
$$+ \operatorname{tr}(Q \Sigma_x) + \operatorname{tr}(RL \Sigma_{\hat{x}_u} L^\intercal)$$
$$+ \operatorname{tr}(S((A+BL) \Sigma_{\hat{x}_u} (A+BL)^\intercal$$
$$+ A \Sigma_{x|\hat{x}_u} A^\intercal + \Sigma_\xi - \Sigma_x))).$$

Here $\beta > 0$ is the Lagrange multiplier corresponding to the constraint $\mathcal{J}_\pi \leq c$ and serving as the marginal trade-off coefficient between the external cost and the information rate, $\tfrac{\beta}{2} S \in \mathbb{R}^{n \times n}$ is the multiplier of the constraint (7) and for convenience the entire Lagrangian is divided by $\beta$. As in rate-distortion theory, $\mathcal{F}$ can be minimized for any given value of $\beta$. The $\beta$ that corresponds to a specific expected cost-rate guarantee level $c$ can then be found using a binary search. The case $\beta = 0$ corresponds to the minimization of information without any cost constraint.

*Theorem 1:* Given $\beta$, the Lagrangian (9) is minimized by a controller satisfying the forward equations

$$\Sigma_x = (A + BL)\Sigma_{\hat{x}_u}(A+BL)^\intercal \qquad (10a)$$
$$\qquad + A\Sigma_{x|\hat{x}_u}A^\intercal + \Sigma_\xi$$
$$\Sigma_y = C\Sigma_x C^\intercal + \Sigma_\epsilon \qquad (10b)$$
$$K = \Sigma_x C^\intercal \Sigma_y^\dagger \qquad (10c)$$
$$\Sigma_{\hat{x}_y} = K\Sigma_y K^\intercal, \qquad (10d)$$

the backward equations

$$M = \beta^{-1} C^\intercal K^\intercal (\Sigma_{\hat{x}_y|\hat{x}_u}^\dagger - \Sigma_{\hat{x}_y}^\dagger)KC \qquad (10e)$$
$$S = Q + A^\intercal S A - M, \qquad (10f)$$
$$L = -(R + B^\intercal S B)^\dagger B^\intercal S A \qquad (10g)$$
$$N = L^\intercal (R + B^\intercal S B) L \qquad (10h)$$

and the control-based estimator covariance

$$\Sigma_{\hat{x}_u} = \Sigma_{\hat{x}_y}^{1/2} V D V^\intercal \Sigma_{\hat{x}_y}^{1/2}, \qquad (10i)$$

the latter determined by the eigenvalue decomposition (EVD)

$$V \Lambda V^\intercal = \Sigma_{\hat{x}_y}^{1/2} N \Sigma_{\hat{x}_y}^{1/2} \qquad (10j)$$

having $V$ orthogonal with $n - \mathrm{rank}(\Sigma_{\hat{x}_y})$ columns spanning the kernel of $\Sigma_{\hat{x}_y}$ and $\Lambda = \mathrm{diag}\{\lambda_i\}$ and by the active mode coefficient matrix

$$D = \mathrm{diag}\left\{\begin{array}{ll} 1 - \beta^{-1}\lambda_i^{-1} & \lambda_i > \beta^{-1} \\ 0 & \lambda_i \leq \beta^{-1} \end{array}\right\}. \qquad (10k)$$

*Proof:* See SM, Appendix IV. $\square$

The spectral analysis in (10j)–(10k) implies that in (10e) the signal-to-noise-ratio (SNR) matrix $Z = \Sigma_{\hat{x}_y|\hat{x}_u}^\dagger - \Sigma_{\hat{x}_y}^\dagger$ satisfies

$$Z = \Sigma_{\hat{x}_y|\hat{x}_u}^\dagger - \Sigma_{\hat{x}_y}^\dagger = \Sigma_{\hat{x}_y}^{\dagger/2} V((I-D)^{-1} - I) V^\intercal \Sigma_{\hat{x}_y}^{\dagger/2}$$
$$= \beta \Sigma_{\hat{x}_y}^{\dagger/2} V D \Lambda V^\intercal \Sigma_{\hat{x}_y}^{\dagger/2}$$

and that the information rate is

$$\mathcal{I}_\pi = \tfrac{1}{2}(\log |\Sigma_{\hat{x}_y}|_\dagger - \log |\Sigma_{\hat{x}_y|\hat{x}_u}|_\dagger) \qquad (11)$$
$$= -\log |I - D| = \sum_i \max(0, \log \beta \lambda_i).$$

As shown in the SM, Appendix I, given an additive Gaussian noise channel $w_t \to \hat{w}_t$ with noise covariance $I - D$, the optimal encoder and decoder are now given by

$$w_t = D^{1/2} V^\intercal \Sigma_{\hat{x}_y}^{\dagger/2} \hat{x}_{y_t}$$
$$\hat{x}_{u_t} = \Sigma_{\hat{x}_y}^{1/2} V D^{1/2} \hat{w}_t,$$

which can be summarized in the form (5b), with

$$W = \Sigma_{\hat{x}_u} \Sigma_{\hat{x}_y}^\dagger$$
$$\Sigma_\omega = \Sigma_{\hat{x}_y}^{1/2} V D(I-D) V^\intercal \Sigma_{\hat{x}_y}^{1/2}.$$

Alternatively, the controller can be given in the form (3), with

$$H = LWK$$
$$\Sigma_\eta = L \Sigma_\omega L^\intercal.$$

Interestingly, Theorem 1 also shows that $S$ corresponds to the cost-to-go Hessian, with respect to the state, as in classic control theory. The difference is that here $S$ also accumulates the non-quadratic information cost and is only the Hessian in an average sense. In the form given in Theorem 1, $M$ is positive semidefinite, but $S$ may not be. This is not problematic if we view $S$ as the Lagrange multiplier of the equality constraint (7), but it is undesired for a cost-to-go Hessian. The positive semidefiniteness of $S$ is discussed further and restored in Part II [15, Section III-C].

Theorem 1 gives the first-order necessary conditions for a solution to be optimal; namely, that the gradient of the Lagrangian (9) is 0 with respect to each parameter. It additionally includes two more conditions, one which is higher-order and the other non-necessary. First, the condition on $\Sigma_{\hat{x}_u}$ (10i) is necessary but not first-order, being a solution to a semidefinite program (see SM, Appendix V). Second, the condition on $L$ (10g) is the least-square solution of a possibly underdetermined system, which means that it may not hold for all optimal solutions but that it does hold for some globally optimal solution.

Problem 1 is highly non-convex and has many local optima that satisfy the first-order necessary conditions. By including the two higher-order and non-necessary conditions, we exclude many of these local optima, although some remain (see Part II [15, Section IV]). This merits further study of the fixed-point structure of this problem.

### B. Phenomenology

To better understand the optimal solution of Theorem 1, consider its phenomenology as $\beta$ spans its range from 0 to $\infty$. The following is the SRD extension of a standard result in one-shot RD theory [37].

*Lemma 3:* Let $\mathcal{I}(\mathcal{J})$ be the minimal information rate achievable by a controller that incurs cost at rate at most $\mathcal{J}$. This information-cost function is monotonically decreasing, convex, and its slope is

$$\partial_\mathcal{J} \mathcal{I} = -\beta, \qquad (12)$$

for $\beta$ the Lagrange multiplier corresponding to the expected cost-rate guarantee level $c = \mathcal{J}$.

*Proof:* For any $\beta$, let

$$\pi^* = \arg\min_\pi \{\beta^{-1} \mathcal{I}_\pi + \mathcal{J}_\pi\}.$$

$\pi^*$ achieves the optimum in Problem 1 when $c = \mathcal{J}_{\pi^*}$. Take

$$\mathcal{I} = \mathcal{I}_{\pi^*}; \qquad \mathcal{J} = \mathcal{J}_{\pi^*}; \qquad \mathcal{F} = \beta^{-1}\mathcal{I} + \mathcal{J}.$$

Then the slope equation follows by fixing $\beta$ while $\mathcal{J}$ and $\mathcal{I}$ vary and noting that at the optimum

$$\partial_\mathcal{J} \mathcal{F} = \beta^{-1} \partial_\mathcal{J} \mathcal{I} + 1 = 0.$$

Monotonicity follows directly from the definition of Problem 1. Convexity can also be shown directly; however, it follows more easily from the slope equation (12) by considering that $\mathcal{J}$ is non-increasing in $\beta$ and thus

$$\partial^2_{\mathcal{J}^2} \mathcal{I} = -\partial_{\mathcal{J}} \beta \geq 0. \qquad \square$$

We now turn to consider how the controller order is increased as $\beta$ is increased from 0 to $\infty$. This phenomenon is known as a water-filling effect [36], [37], and is made explicit in the form of the optimal information rate $\mathcal{I}_\pi$ (11). Note, however, that in the SRD problem the water-filling effect is self-consistent, in that $\Lambda$ itself depends on $\beta$.

*Definition 5:* The order of a LTI controller is $\mathrm{rank}(\Sigma_{\hat{x}_u})$. For the optimal solution (10i), this equals $\mathrm{rank}(D)$, the number of active modes.

Let us consider a stable plant, having all eigenvalues of $A$ inside the unit circle. We note that our results hold more generally and extend known results [18] when the plant is unstable but stabilizable and detectable. However, the analysis of this case when $\beta \to 0$ is more involved and is presented separately in an upcoming paper.

When $\beta = 0$, we are only interested in minimizing $\mathcal{I}_\pi$ and therefore take an order-0 controller, having $D = 0$, $\Sigma_{\hat{x}_u} = 0$ and $M = 0$. $\Sigma_x$ and $S$ satisfy the uncontrolled Lyapunov equations

$$\Sigma_x = A\,\Sigma_x\,A + \Sigma_\xi$$
$$S = Q + A^\intercal SA.$$

$L$ and $N$ can be set accordingly, despite the fact that no attention to the observation is spent and no control is possible. Computing the EVD of $\Sigma_{\hat{x}_y}$ and applying (10j), we can retrieve $\Lambda$.

As we increase $\beta$, this uncontrolled solution remains constant as long as $\beta \leq \lambda_1^{-1}$, the inverse of the largest eigenvalue in $\Lambda$. At that first critical point, the controller undergoes a phase transition, where its order increases from 0 to 1 (or higher if $\lambda_1$ is not unique in $\Lambda$).

Note that $\Lambda$ contains the same eigenvalues as the matrix

$$\Sigma_{N^{1/2}\hat{x}_y} = N^{1/2}\,\Sigma_{\hat{x}_y}\,N^{1/2},$$

which represents the value of the information that the observation has on the state, in terms of the cost reduction it allows. Thus an order-1 controller observes and controls the state mode that provides the largest decrease in cost per bit of observed information, in keeping with (12).

Beyond the first phase transition, the optimal solution does change with $\beta$ and so does $\Lambda$. Eventually, $\beta$ meets $\lambda_i^{-1}(\beta)$, for each $i = 2, \ldots, \mathrm{rank}(\Sigma_{N^{1/2}\hat{x}_y})$ in turn and further phase transitions occur, increasing the controller order until it reaches $\mathrm{rank}(\Sigma_{N^{1/2}\hat{x}_y})$.

As long as $\beta$ is finite, even after the last phase transition, the information rate must be finite. Since the controller lacks the capacity to attend to any mode with perfect fidelity, it must maintain some uncertainty in all modes and accordingly $D \prec I$ and $\Sigma_{\hat{x}_u} \prec \Sigma_{\hat{x}_y}$. As $\beta \to \infty$, the SNR matrix $Z = \Sigma^\dagger_{\hat{x}_y|\hat{x}_u} - \Sigma^\dagger_{\hat{x}_y}$ grows to infinity in modes having $\lambda_i > 0$, as does the information rate in these modes.

The $\beta = \infty$ case marks a qualitative change in the optimization problem. We are no longer concerned with the information rate and only wish to minimize the expected cost rate $\mathcal{J}_\pi$. The optimal solution here is underdetermined with respect to useless modes where $\lambda_i = 0$. Despite having no value in decreasing $\mathcal{J}_\pi$, at $\beta = \infty$ (but not for $\beta \to \infty$) these modes may be observed for free. This allows us to simplify the solution to

$$D = I$$
$$\Sigma_{\hat{x}_u} = \Sigma_{\hat{x}_y}$$
$$M = C^\intercal K^\intercal \Sigma^{\dagger/2}_{\hat{x}_y} \Sigma^{1/2}_{\hat{x}_y} N \Sigma^{1/2}_{\hat{x}_y} \Sigma^{\dagger/2}_{\hat{x}_y} KC$$
$$= C^\intercal K^\intercal NKC.$$

It is interesting to note the impact of the observability on $M$ at $\beta = \infty$. When the plant is unobservable, we have $C = K = 0$ and thus $M = 0$. When observability is full, we have $C = K = I$ and thus $M = N$. For partial observability models, $N - M$ is not necessarily positive semidefinite, which will become important in the reduced retentive control problem (see Part II [15, Section III-C]).

In the classic control problem, where observability is partial but the memory and the sensory capacities are unbounded, the memory state is maintained by the Kalman filter and we have $M = N$ and

$$S = Q + A^\intercal SA - N,$$

independent of the forward inference process. Note, however, that $S$ in that case is the Hessian of the certainty-equivalent cost-to-go with respect to $\hat{x}_t$, instead of $x_t$.

Thus either full and unbounded ($\beta = \infty$) observability or bounded ($\beta < \infty$) sensing with unbounded memory [31] are sufficient for recovering the separation principle of classic control theory. In the more general case, the backward control process (10f) is coupled with the forward inference process (10a).

## V. DISCUSSION

In this paper we introduce the problem of optimal memoryless LQG control with bounded channel capacity. We present the solution and discuss some of its properties and phenomenology.

Part of our motivation in considering memoryless controllers is that the problem of retentive (memory-utilizing) control can be reduced to the problem of memoryless control. This is discussed in detail in Part II of this work [15, Section III-B]. The two control models are also compared there (Section IV) using an illustrative example.

One attractive aspect of our solution is its principled reduction of the controller order. In many applications, the controller's information rate is a more natural measure of its complexity than the dimension of its support. Nevertheless, a hard constraint on the order is sometimes required, alongside a soft constraint on the information rate, leading to an algorithmically challenging open question.

The controllers considered in this paper have linear-Gaussian control laws. This class of controllers does not solve optimally all control problems and is particularly prone to suboptimality in memory-constrained settings [38], [39]. Nevertheless, we conjecture that there exist some moderately strong conditions under which the bounded memoryless control problem discussed here is solved optimally by an LTI controller.

# Minimum-Information LQG Control
# Supplementary Material

Roy Fox[†] and Naftali Tishby[†]

## APPENDIX I
## PERFECTLY MATCHED CHANNEL

In this appendix we construct a channel that is perfectly matched to the sequential source code derived in Theorem 1, in Part I of this paper [1, Section III-B]. Recall that in a perfectly matched source-channel pair the optimal source coding and the optimal channel coding can be implemented jointly for single letters, without requiring longer blocks. This allows us to use them in a perception-action cycle, where we cannot accumulate a block of inputs before emitting an output.

The main results of [2], applied to our setting, can be summarized as follows. We wish to find a memoryless channel into which we can input an encoding $w_t = g(\hat{x}_{y_t})$, such that $\hat{x}_{u_t} = h(\hat{w}_t)$ can be decoded from the channel output $\hat{w}_t$. Suppose that we are concerned with the power needed to transmit $w_t$ and thus the input cost is $w_t^\intercal w_t$. Then the source $\hat{x}_{y_t}$ and the channel $w_t \to \hat{w}_t$ are perfectly matched if there exist an encoder and a decoder such that

1) The Kullback-Leibler divergence $\mathbb{D}[f(\hat{w}_t|w_t)\|f(\hat{w}_t)]$ between the conditional and marginal densities of $\hat{w}_t$, as a function of $w_t$, equals $c_1 w_t^\intercal w_t + c_2$, for some constants $c_1 \geq 0$ and $c_2$; and
2) $f(\hat{x}_{u_t}|\hat{x}_{y_t})$ satisfies the conditions in Theorem 1.

To meet these conditions, we can choose the channel, the encoder and the decoder to have

$$w_t = D^{1/2} V^\intercal \Sigma_{\hat{x}_y}^{\dagger/2} \hat{x}_{y_t}$$
$$\hat{w}_t = w_t + v_t; \qquad v_t \sim \mathcal{N}(0, I - D)$$
$$\hat{x}_{u_t} = \Sigma_{\hat{x}_y}^{1/2} V D^{1/2} \hat{w}_t,$$

with $D$ and $V$ as in Theorem 1. Then

$$\Sigma_w = D$$
$$\Sigma_{\hat{w}} = I$$
$$\Sigma_{\hat{x}_u} = \Sigma_{\hat{x}_y}^{1/2} V D V^\intercal \Sigma_{\hat{x}_y}^{1/2} = \Sigma_{\hat{x}_u;\hat{x}_y},$$

and it can be verified that

$$\mathbb{D}[f(\hat{w}_t|w_t)\|f(\hat{w}_t)] = \tfrac{1}{2} w_t^\intercal \Sigma_{\hat{w}}^{-1} w_t + \text{const},$$

as required.

The capacity of the additive Gaussian noise channel with noise covariance $I - D$, under the appropriate expected power constraint, is indeed achieved by a Gaussian input with covariance $D$ and is equal to the information rate in Theorem 1. As shown in [2], this means that constraining the expected power $\Sigma_w$ is equivalent to constraining the information rate $\mathbb{I}[\hat{x}_{y_t}; \hat{x}_{u_t}]$.

Note, however, that the matched channel noise covariance depends on the constraint, through the solution in Theorem 1. Moreover, this result is not applicable when the best channel available to the designer of the controller is not the matched channel above, in which case both the channel and the sequential source coding generally need to be adapted.

## APPENDIX II
## PROOF OF LEMMA 1 OF PART I

In this appendix we restate and prove Lemma 1 of Part I [1, Section IV-A].

*Lemma 1:* Let $x$ and $\hat{x}$ be 0-mean jointly Gaussian random variables. The following properties are equivalent:

1) There exists a random variable $u$, jointly Gaussian with $x$, such that $\hat{x}(u) = \arg\min_{\hat{x}} \mathbb{E}[\|\hat{x} - x\|^2 | u] = \mathbb{E}[x|u]$.
2) $\Sigma_{\hat{x};x} = \Sigma_{\hat{x}}$.
3) $\Sigma_{x|\hat{x}} = \Sigma_x - \Sigma_{\hat{x}}$, where $\Sigma_{x|\hat{x}}$ is the conditional covariance matrix of $x$ given $\hat{x}$, implying $\Sigma_x \succeq \Sigma_{\hat{x}}$.
4) $\hat{x} = \mathbb{E}[x|\hat{x}]$.

Such $\hat{x}$ is called a minimum mean square error (MMSE) estimator (of $u$) for $x$.

*Proof:* (1 $\implies$ 2) Assume without loss of generality that $u$ has mean 0. Then

$$\hat{x} = \Sigma_{x;u} \Sigma_u^\dagger u,$$

implying

$$\Sigma_{\hat{x};x} = \Sigma_{x;u} \Sigma_u^\dagger \Sigma_{u;x} = \Sigma_{\hat{x}}.$$

(2 $\implies$ 3)

$$\Sigma_{x|\hat{x}} = \Sigma_x - \Sigma_{x;\hat{x}} \Sigma_{\hat{x}}^\dagger \Sigma_{\hat{x};x} = \Sigma_x - \Sigma_{\hat{x}}.$$

(3 $\implies$ 4) Since $x$ and $\hat{x}$ are 0-mean and jointly Gaussian, we can write for some $T$

$$x = T\hat{x} + \xi; \qquad \xi \sim \mathcal{N}(0, \Sigma_{x|\hat{x}}),$$

implying

$$\Sigma_x = T \Sigma_{\hat{x}} T^\intercal + \Sigma_x - \Sigma_{\hat{x}},$$

thus without loss of generality $T = I$.

[†]School of Computer Science and Engineering, The Hebrew University, {royf,tishby}@cs.huji.ac.il
[*]This work was supported by the DARPA MSEE Program, the Gatsby Charitable Foundation, the Israel Science Foundation and the Intel ICRI-CI Institute

($4 \implies 1$) Taking $u = \hat{x}$, we have

$$\arg\min_{\hat{x}'} \mathbb{E}[\|\hat{x}' - x\|^2 | u]$$
$$= \arg\min_{\hat{x}'}(\hat{x}'^\mathsf{T}\hat{x}' - 2\hat{x}'^\mathsf{T}\,\mathbb{E}[x|u]) + \mathbb{E}[x^\mathsf{T} x|u],$$

which is optimized by $\hat{x}' = \mathbb{E}[x|u]$. $\square$

## APPENDIX III
### PROOF OF LEMMA 2 OF PART I

In this appendix we restate and prove Lemma 2 of Part I [1, Section IV-A].

*Lemma 2:* The bounded memoryless LTI controller optimization problem (Problem 1) is solved by a control law of the form

$$\hat{x}_{y_t} = K y_t \tag{5a}$$
$$\hat{x}_{u_t} = W\hat{x}_{y_t} + \omega_t; \qquad \omega_t \sim \mathcal{N}(0, \Sigma_\omega) \tag{5b}$$
$$u_t = L\hat{x}_{u_t}, \tag{5c}$$

where $W \in \mathbb{R}^{n \times n}$, $\Sigma_\omega \in \mathbb{R}^{n \times n}$, $L \in \mathbb{R}^{\ell \times n}$, $\omega_t$ is independent of $y_t$, $\hat{x}_{u_t}$ is a MMSE estimator for $\hat{x}_{y_t}$ and

$$\mathbb{I}[y_t; u_t] = \mathbb{I}[\hat{x}_{y_t}; \hat{x}_{u_t}]. \tag{6}$$

*Proof:* Consider a LTI controller $\pi$ of the form

$$u_t = H y_t + \eta_t; \qquad \eta_t \sim \mathcal{N}(0, \Sigma_\eta), \tag{III.1}$$

satisfying the Markov network

$$
\begin{array}{ccc}
x_t \text{ --- } y_t & \text{ --- } & u_t \\
| & & | \\
\hat{x}_{y_t} & & \hat{x}_{u_t}.
\end{array}
$$

We now construct a controller $\pi'$ with control law $u'_t$ based on the estimator $\hat{x}'_{u_t}$ by defining the Markov chain

$$x_t \text{ --- } y_t \text{ --- } \hat{x}_{y_t} \text{ --- } u''_t \text{ --- } \hat{x}'_{u_t} \text{ --- } u'_t$$

such that each consecutive pair of variables has the same joint distribution as their unprimed namesakes. Since $\hat{x}_{y_t}$ is a sufficient statistic of $y_t$ for $x_t$, we have the Markov chain $x_t \text{ --- } \hat{x}_{y_t} \text{ --- } y_t \text{ --- } u_t$, implying that $u''_t$ has the same joint distribution with $x_t$ as $u_t$ does. Likewise, $\hat{x}'_{u_t}$ has the same joint distribution with $x_t$ as $\hat{x}_{u_t}$ does. Since $\hat{x}_{u_t}$ is a sufficient statistic of $u_t$ for $x_t$, we have that $u'_t$ also has the same joint distribution with $x_t$ as $u_t$ does.

Thus the controller $\pi'$ induces the same stochastic process $\{x_t, u'_t\}$ and the same external cost. Note that $u'_t$ may not have the same joint distribution with $y_t$ as $u_t$ does and due to the data-processing inequality [3]

$$\mathbb{I}[y_t; u_t] \geq \mathbb{I}[\hat{x}_{y_t}; u_t] = \mathbb{I}[\hat{x}_{y_t}; u''_t]$$
$$\geq \mathbb{I}[\hat{x}_{y_t}; \hat{x}'_{u_t}] \geq \mathbb{I}[y_t; u'_t].$$

Therefore $\pi'$ performs at least as well as $\pi$ and equally well when $\pi$ is optimal, proving (6).

$\hat{x}'_{u_t}$ is a MMSE estimator for $\hat{x}_{y_t}$ since

$$\mathbb{E}[\hat{x}_{y_t}|\hat{x}'_{u_t}] = \mathbb{E}[\mathbb{E}[x_t|y_t]|\hat{x}_{u'_t}]$$
$$= \mathbb{E}[x_t|\hat{x}'_{u_t}] = \hat{x}'_{u_t},$$

where the second equality follows from $x_t \text{ --- } y_t \text{ --- } \hat{x}'_{u_t}$.

Finally, it may not be clear from the above analysis that $u'_t$ is optimally deterministic in $\hat{x}'_{u_t}$. If $u_t$ has covariance $\Sigma_\nu$ given $\hat{x}'_{u_t}$, the Lagrangian of the optimization problem ((9) in Part I) depends on $\Sigma_\nu$ only through the terms

$$\tfrac{1}{2}(\mathrm{tr}(R\,\Sigma_\nu) + \mathrm{tr}(SB\,\Sigma_\nu\,B^\mathsf{T})).$$

Since $R + B^\mathsf{T} S B \succeq 0$ is positive semidefinite, we can take $\Sigma_\nu = 0$ without loss of performance, recovering the structure (5). Intuitively, the argument is that any noise added to $u'_t$, beyond $\hat{x}'_{u_t}$, is not helpful in compressing $x_t$ and can only increase the external cost without saving any communication cost.

In the other direction, let $u_t$ satisfy the form of Lemma 2. We can rewrite $u_t$ in the form (III.1), with

$$H = LWK$$
$$\Sigma_\eta = L\,\Sigma_\omega\,L^\mathsf{T}. \qquad \square$$

## APPENDIX IV
### PROOF OF THEOREM 1 OF PART I

In this appendix we restate and prove Theorem 1 of Part I [1, Section IV-A], which relies on the following Lagrangian developed there.

$$\mathcal{F}_{\Sigma_x, \Sigma_{\hat{x}_u}, L, S; \beta} = \tfrac{1}{2}(\beta^{-1}(\log |\Sigma_{\hat{x}_y}|_\dagger - \log |\Sigma_{\hat{x}_y | \hat{x}_u}|_\dagger) \tag{9}$$
$$+ \mathrm{tr}(Q\,\Sigma_x) + \mathrm{tr}(RL\,\Sigma_{\hat{x}_u}\,L^\mathsf{T})$$
$$+ \mathrm{tr}(S((A+BL)\,\Sigma_{\hat{x}_u}\,(A+BL)^\mathsf{T}$$
$$+ A\,\Sigma_{x|\hat{x}_u}\,A^\mathsf{T} + \Sigma_\xi - \Sigma_x))).$$

*Theorem 1:* Given $\beta$, the Lagrangian (9) is minimized by a controller satisfying the forward equations

$$\Sigma_x = (A+BL)\,\Sigma_{\hat{x}_u}(A+BL)^\mathsf{T} \tag{10a}$$
$$\qquad + A\,\Sigma_{x|\hat{x}_u}\,A^\mathsf{T} + \Sigma_\xi$$
$$\Sigma_y = C\,\Sigma_x\,C^\mathsf{T} + \Sigma_\epsilon \tag{10b}$$
$$K = \Sigma_x\,C^\mathsf{T}\,\Sigma_y^\dagger \tag{10c}$$
$$\Sigma_{\hat{x}_y} = K\,\Sigma_y\,K^\mathsf{T}, \tag{10d}$$

the backward equations

$$M = \beta^{-1} C^\mathsf{T} K^\mathsf{T} (\Sigma_{\hat{x}_y | \hat{x}_u}^\dagger - \Sigma_{\hat{x}_y}^\dagger) KC \tag{10e}$$
$$S = Q + A^\mathsf{T} SA - M, \tag{10f}$$
$$L = -(R + B^\mathsf{T} SB)^\dagger B^\mathsf{T} SA \tag{10g}$$
$$N = L^\mathsf{T}(R + B^\mathsf{T} SB)L \tag{10h}$$

and the control-based estimator covariance

$$\Sigma_{\hat{x}_u} = \Sigma_{\hat{x}_y}^{1/2} V D V^\mathsf{T} \Sigma_{\hat{x}_y}^{1/2}, \tag{10i}$$

the latter determined by the eigenvalue decomposition (EVD)

$$V \Lambda V^\mathsf{T} = \Sigma_{\hat{x}_y}^{1/2} N \Sigma_{\hat{x}_y}^{1/2} \tag{10j}$$

having $V$ orthogonal with $n - \mathrm{rank}(\Sigma_{\hat{x}_y})$ columns spanning the kernel of $\Sigma_{\hat{x}_y}$ and $\Lambda = \mathrm{diag}\{\lambda_i\}$ and by the active mode coefficient matrix

$$D = \mathrm{diag}\left\{ \begin{array}{ll} 1 - \beta^{-1}\lambda_i^{-1} & \lambda_i > \beta^{-1} \\ 0 & \lambda_i \leq \beta^{-1} \end{array} \right\}. \tag{10k}$$

*Proof:* The minimum of the Lagrangian (9) must satisfy the first-order optimality conditions, i.e. that the gradient with respect to each parameter is 0 at the optimum. We start by differentiating $\mathcal{F}$ by the feedback gain $L$

$$\partial_L \mathcal{F}_{\Sigma_x, \Sigma_{\hat{x}_u}, L, S; \beta} = RL\Sigma_{\hat{x}_u} + B^\intercal S(A + BL)\Sigma_{\hat{x}_u} = 0,$$

which we rewrite as

$$(R + B^\intercal SB)L\Sigma_{\hat{x}_u} = -B^\intercal SA\Sigma_{\hat{x}_u}.$$

As this equation shows, $L$ is underdetermined in the kernel of $\Sigma_{\hat{x}_u}$, since these modes are always 0 in $\hat{x}_{u_t}$ and have no effect on $u_t$. $L$ is also underdetermined in the kernel of $R + B^\intercal SB$, since these modes have no cost (immediate or future) and can be controlled in any way without affecting the solution's performance. Thus without loss of performance we can take

$$L = -(R + B^\intercal SB)^\dagger B^\intercal SA.$$

We substitute this solution back into the Lagrangian, to get

$$\mathcal{F}_{\Sigma_x, \Sigma_{\hat{x}_u}, S; \beta} = \tfrac{1}{2}(\beta^{-1}(\log|\Sigma_{\hat{x}_y}|_\dagger - \log|\Sigma_{\hat{x}_y|\hat{x}_u}|_\dagger) \quad \text{(IV.1)}$$
$$+ \operatorname{tr}(M\Sigma_x) - \operatorname{tr}(N\Sigma_{\hat{x}_u}) + \operatorname{tr}(S\Sigma_\xi)),$$

with

$$M = Q + A^\intercal SA - S$$
$$N = L^\intercal (R + B^\intercal SB)L$$
$$= A^\intercal SB(R + B^\intercal SB)^\dagger B^\intercal SA.$$

The problem of optimizing over $\Sigma_{\hat{x}_u}$ given the other parameters can now be written, up to constants, as the semidefinite program (SDP)

$$\max_{\Sigma_{\hat{x}_u}} \quad \log|\Sigma_{\hat{x}_y} - \Sigma_{\hat{x}_u}|_\dagger + \beta \operatorname{tr}(N\Sigma_{\hat{x}_u})$$
$$\text{s.t.} \quad 0 \preceq \Sigma_{\hat{x}_u} \preceq \Sigma_{\hat{x}_y}.$$

By Lemma V.1 in Appendix V, the optimum is achieved when $\Sigma_{\hat{x}_u}$ satisfies (10i)–(10k).

Finally, with $P = \Sigma_{\hat{x}_y}\Sigma_{\hat{x}_y}^\dagger$ the projection onto the support of $\hat{x}_{y_t}$ and since the range of $\Sigma_{\hat{x}_u}$ is contained in that subspace, we have

$$\partial_{(\Sigma_x)_{i,j}}(\log|\Sigma_{\hat{x}_y}|_\dagger - \log|\Sigma_{\hat{x}_y|\hat{x}_u}|_\dagger)$$
$$= -\partial_{(\Sigma_x)_{i,j}} \log|P - \Sigma_{\hat{x}_u}\Sigma_{\hat{x}_y}^\dagger|_\dagger$$
$$= -\partial_{(\Sigma_x)_{i,j}} \log|I - \Sigma_{\hat{x}_u}(P\Sigma_{\hat{x}_y}P)^\dagger|$$
$$= \operatorname{tr}((I - \Sigma_{\hat{x}_u}\Sigma_{\hat{x}_y}^\dagger)^{-1}\Sigma_{\hat{x}_u}\partial_{(\Sigma_x)_{i,j}}(P\Sigma_{\hat{x}_y}P)^\dagger).$$

The purpose of introducing $P$ is to notice that even if the range of $\Sigma_{\hat{x}_y}$ is increased, this has no effect on the Lagrangian, because these modes are orthogonal to the range of $\Sigma_{\hat{x}_u}$. This allows us to treat $P$ as constant, so that the range of $P\Sigma_{\hat{x}_y}P$ is constant in a neighborhood of the solution, and the derivative of the pseudoinverse is simplified in this case to

$$\partial_{(\Sigma_x)_{i,j}}(P\Sigma_{\hat{x}_y}P)^\dagger = -\Sigma_{\hat{x}_y}^\dagger(\partial_{(\Sigma_x)_{i,j}}\Sigma_{\hat{x}_y})\Sigma_{\hat{x}_y}^\dagger$$
$$= -\Sigma_{\hat{x}_y}^\dagger KCJ_{i,j}C^\intercal K^\intercal \Sigma_{\hat{x}_y}^\dagger,$$

with $J_{i,j}$ the matrix with 1 in position $(i,j)$ and 0 elsewhere. This yields

$$\partial_{\Sigma_x}\mathcal{F}_{\Sigma_x,\Sigma_{\hat{x}_u},S;\beta}$$
$$= \tfrac{1}{2}(M - \beta^{-1}C^\intercal K^\intercal \Sigma_{\hat{x}_y}^\dagger(I - \Sigma_{\hat{x}_u}\Sigma_{\hat{x}_y}^\dagger)^{-1}\Sigma_{\hat{x}_u}\Sigma_{\hat{x}_y}^\dagger KC)$$
$$= \tfrac{1}{2}(M - \beta^{-1}C^\intercal K^\intercal \Sigma_{\hat{x}_y}^\dagger((I - \Sigma_{\hat{x}_u}\Sigma_{\hat{x}_y}^\dagger)^{-1} - I)KC)$$
$$= \tfrac{1}{2}(M - \beta^{-1}C^\intercal K^\intercal (\Sigma_{\hat{x}_y|\hat{x}_u}^\dagger - \Sigma_{\hat{x}_y}^\dagger)KC) = 0,$$

implying (10e). □

## APPENDIX V
## SEMIDEFINITE PROGRAM SOLUTION

In this appendix we state and prove the following solution to our SDP problem.

*Lemma V.1:* The semidefinite program

$$\max_{X \in \mathbb{S}_+^n} \quad \log|M_1 - X|_\dagger + \operatorname{tr}(M_2 X)$$
$$\text{s.t.} \quad X \preceq M_1,$$

with $M_1, M_2 \succeq 0$, is optimized by

$$X = M_1^{1/2} VDV^\intercal M_1^{1/2},$$

with the eigenvalue decomposition (EVD)

$$V\Lambda V^\intercal = M_1^{1/2} M_2 M_1^{1/2},$$

such that $V$ is orthogonal with $n - \operatorname{rank}(M_1)$ columns spanning the kernel of $M_1$ and $\Lambda = \operatorname{diag}\{\lambda_i\}$ and with

$$D = \operatorname{diag}\left\{\begin{array}{ll} 1 - \lambda_i^{-1} & \lambda_i > 1 \\ 0 & \lambda_i \leq 1 \end{array}\right\}.$$

*Proof:* Let the EVD of $M_1$ be

$$U\Psi U^\intercal = M_1,$$

with $U$ orthogonal and $\Psi$ diagonal, having

$$\Psi = \begin{bmatrix} \Psi_+ & 0 \\ 0 & 0_{(n-m)\times(n-m)} \end{bmatrix},$$

with $m = \operatorname{rank}(M_1)$. Let

$$\Psi^\ddagger = \Psi^\dagger + I - \Psi^\dagger\Psi = \begin{bmatrix} \Psi_+^{-1} & 0 \\ 0 & I \end{bmatrix}.$$

By changing the variable to

$$Y = \Psi^{\ddagger/2} U^\intercal X U \Psi^{\ddagger/2},$$

the constraint of the SDP becomes

$$Y \preceq I_{m,n} = \begin{bmatrix} I_{m\times m} & 0 \\ 0 & 0_{(n-m)\times(n-m)} \end{bmatrix}.$$

$Y$ must therefore be 0 outside the upper-left $m \times m$ block, and the SDP is equivalent, up to constants, to

$$\max_{Y \in \mathbb{S}_+^n} \quad \log|I_{m,n} - Y|_\dagger + \operatorname{tr}(\Psi^{1/2}U^\intercal M_2 U\Psi^{1/2}Y)$$
$$\text{s.t.} \quad Y \preceq I_{m,n}.$$

Let the EVD of the linear coefficient be

$$\bar{V}\Lambda\bar{V}^\intercal = \Psi^{1/2}U^\intercal M_2 U\Psi^{1/2},$$

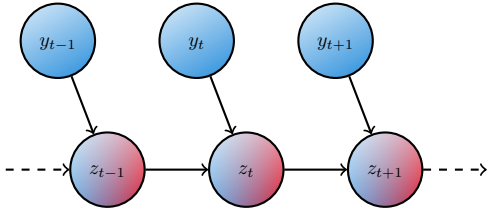

Fig. VI.1. Bayesian network of online inference from a sequence of independent observations

with

$$\bar{V} = \begin{bmatrix} \bar{V}_+ & 0 \\ 0 & I_{(n-m)\times(n-m)} \end{bmatrix}$$

orthogonal and preserving the kernel of $\Psi$ and $\Lambda = \text{diag}\{\lambda_i\}$. We can again change the variable to

$$D = \bar{V}^\intercal Y \bar{V},$$

to get

$$\max_{D \in \mathbb{S}_+^n} \log |I_{m,n} - D|_\dagger + \text{tr}(\Lambda D)$$
$$\text{s.t.} \quad D \preceq I_{m,n},$$

which can easily be solved using Hadamard's inequality [3], to find

$$D = \text{diag}\left\{\begin{matrix} 1 - \lambda_i^{-1} & \lambda_i > 1 \\ 0 & \lambda_i \leq 1 \end{matrix}\right\}.$$

Finally, the lemma follows by unmaking the variable changes and taking

$$V = U\bar{V}. \qquad \square$$

## APPENDIX VI
### PROPERTIES OF THE RETENTIVE DIRECTED INFORMATION

In this appendix we show how the retentive directed information (Definition 6 of Part II [4, Section III-A]) relates to the multi-information of Bayesian networks [5].

Consider the Bayesian network in Figure VI.1, which describes the process of online inference from a sequence of independent observations. The multi-information of this network, for horizon $T$, is equal to the retentive directed information

$$\mathbb{I}[y^T, z^T] = \mathbb{E}\left[\log \frac{f(y^T, z^T)}{\prod_{t=1}^T f(y_t)f(z_t)}\right]$$
$$= \sum_{t=1}^T \mathbb{E}\left[\log \frac{f(z_t|z^{t-1}, y^t)}{f(z_t)}\right] = \mathbb{I}[y^T \twoheadrightarrow z^T].$$

An important property of the directed information is that the mutual information between two sequences can be decomposed into the sum of directed information in both directions [6]

$$\mathbb{I}[x^T; z^T] = \mathbb{I}[x^T \to z^T] + \mathbb{I}[z^T \to x^T].$$

Interestingly, retentive directed information extends this property to the retentive control process (Figure 1 in Part II). This process can be thought of as consisting of four phases: observation, inference, control and state transition. Its multi-information can accordingly be decomposed [7] into the sum

$$\mathbb{I}[x^T, y^T, z^T, u^T] = \mathbb{I}[x^T \twoheadrightarrow y^T] + \mathbb{I}[y^T \twoheadrightarrow z^T]$$
$$+ \mathbb{I}[z^T \twoheadrightarrow u^T] + \mathbb{I}[u^T \twoheadrightarrow x^T].$$

## APPENDIX VII
### STRUCTURE OF THE OPTIMAL RETENTIVE CONTROLLER

In this appendix we derive the structure of the optimal retentive controller summarized in Part II [4, Section III-C].

For the structured feedback gain $L$ we find using the Schur complement that

$$(R + B^\intercal SB)^\dagger = \begin{bmatrix} R_u + B_{x;u}^\intercal S_x B_{x;u} & B_{x;u}^\intercal S_{x;m} \\ S_{m;x} B_{x;u} & S_m \end{bmatrix}^\dagger$$
$$= \begin{bmatrix} S_{u|m}^\dagger & -S_{u|m}^\dagger B_{x;u}^\intercal S_{x;m} S_m^\dagger \\ -S_m^\dagger S_{m;x} B_{x;u} S_{u|m}^\dagger & S_{m|u}^\dagger \end{bmatrix},$$

with

$$S_{m|u}^\dagger = S_m^\dagger + S_m^\dagger S_{m;x} B_{x;u} S_{u|m}^\dagger B_{x;u}^\intercal S_{x;m} S_m^\dagger,$$

and so

$$L = -(R + B^\intercal SB)^\dagger B^\intercal SA$$
$$= -(R + B^\intercal SB)^\dagger \begin{bmatrix} B_{x;u}^\intercal S_x A_x & 0 \\ S_{m;x} A_x & 0 \end{bmatrix}$$
$$= -\begin{bmatrix} S_{u|m}^\dagger B_{x;u}^\intercal S_{x|m} A_x & 0 \\ S_m^\dagger S_{m;x}(I - B_{x;u} S_{u|m}^\dagger B_{x;u}^\intercal S_{x|m})A_x & 0 \end{bmatrix}$$
$$= \begin{bmatrix} L_{u;x|m} & 0 \\ -S_m^\dagger S_{m;x}(A_x + B_{x;u} L_{u;x|m}) & 0 \end{bmatrix},$$

with

$$L_{u;x|m} = -S_{u|m}^\dagger B_{x;u}^\intercal S_{x|m} A_x.$$

We also have

$$N = L^\intercal (R + B^\intercal SB) L$$
$$= A^\intercal SB(R + B^\intercal SB)^\dagger B^\intercal SA = \begin{bmatrix} N_{x|m} & 0 \\ 0 & 0 \end{bmatrix}$$
$$N_{x|m} = \begin{bmatrix} B_{x;u}^\intercal S_x A_x \\ S_{m;x} A_x \end{bmatrix}^\intercal L \begin{bmatrix} I \\ 0 \end{bmatrix}$$
$$= A_x^\intercal (S_x - S_{x|m} + S_{x|m} B_{x;u} S_{u|m}^\dagger B_{x;u}^\intercal S_{x|m}) A_x.$$

Dually, for the structured Kalman gain $K$ we find that

$$\Sigma_{\tilde{y}}^\dagger = \begin{bmatrix} \Sigma_y & \Sigma_{y;m} \\ \Sigma_{m;y} & \Sigma_m \end{bmatrix}^\dagger$$
$$= \begin{bmatrix} \Sigma_{y|m}^\dagger & -\Sigma_{y|m}^\dagger \Sigma_{y;m} \Sigma_m^\dagger \\ -\Sigma_m^\dagger \Sigma_{m;y} \Sigma_{y|m}^\dagger & \Sigma_m^\dagger + \Sigma_m^\dagger \Sigma_{m;y} \Sigma_{y|m}^\dagger \Sigma_{y;m} \Sigma_m^\dagger \end{bmatrix},$$

and so

$$K = \Sigma_x C^\intercal \Sigma_{\tilde{y}}^\dagger$$
$$= \begin{bmatrix} \Sigma_x C_{y;x}^\intercal & \Sigma_{x;m} \end{bmatrix} \begin{bmatrix} \Sigma_y & \Sigma_{y;m} \\ \Sigma_{m;y} & \Sigma_m \end{bmatrix}^\dagger$$
$$= \begin{bmatrix} K_{x;y|m} & (I - K_{x;y|m} C_{y;x}) \Sigma_{x;m} \Sigma_m^\dagger \end{bmatrix},$$

with
$$K_{x;y|m} = \Sigma_{x|m} C_{y;x}^\intercal \Sigma_{y|m}^\dagger.$$

Now constraining the controller to be MMSE, we have the structure
$$\Sigma_{\tilde{x}} = \begin{bmatrix} \Sigma_{x|m} + \Sigma_m & \Sigma_m \\ \Sigma_m & \Sigma_m \end{bmatrix}$$
$$K = \begin{bmatrix} K_{x;y|m} & I - K_{x;y|m} C_{y;x} \end{bmatrix},$$

which we employ in differentiating $\mathcal{F}$ (IV.1), to get

$$\partial_{\Sigma_{x|m}} \mathcal{F}_{\Sigma_{x|m}, \Sigma_m, \Sigma_{\hat{x}_{\tilde{u}}}, S; \beta} = \begin{bmatrix} I \\ 0 \end{bmatrix}^\intercal \partial_{\Sigma_{\tilde{x}}} \mathcal{F}_{\Sigma_{\tilde{x}}, \Sigma_{\hat{x}_{\tilde{u}}}, S; \beta} \begin{bmatrix} I \\ 0 \end{bmatrix}$$
$$= \tfrac{1}{2} \begin{bmatrix} I \\ 0 \end{bmatrix}^\intercal (M - \beta^{-1} C^\intercal K^\intercal Z K C) \begin{bmatrix} I \\ 0 \end{bmatrix}$$
$$= \tfrac{1}{2} \left( \begin{bmatrix} I \\ 0 \end{bmatrix}^\intercal M \begin{bmatrix} I \\ 0 \end{bmatrix} - \beta^{-1} C_{y;x}^\intercal K_{x;y|m}^\intercal Z K_{x;y|m} C_{y;x} \right) = 0$$
$$\partial_{\Sigma_m} \mathcal{F}_{\Sigma_{x|m}, \Sigma_m, \Sigma_{\hat{x}_{\tilde{u}}}, S; \beta} = \begin{bmatrix} I \\ I \end{bmatrix}^\intercal \partial_{\Sigma_{\tilde{x}}} \mathcal{F}_{\Sigma_{\tilde{x}}, \Sigma_{\hat{x}_{\tilde{u}}}, S; \beta} \begin{bmatrix} I \\ I \end{bmatrix}$$
$$= \tfrac{1}{2} \begin{bmatrix} I \\ I \end{bmatrix}^\intercal (M - \beta^{-1} C^\intercal K^\intercal Z K C) \begin{bmatrix} I \\ I \end{bmatrix}$$
$$= \tfrac{1}{2} \left( \begin{bmatrix} I \\ I \end{bmatrix}^\intercal M \begin{bmatrix} I \\ I \end{bmatrix} - \beta^{-1} Z \right) = 0,$$

with
$$Z = \Sigma_{\hat{x}_{\tilde{y}}|\hat{x}_{\tilde{u}}}^\dagger - \Sigma_{\hat{x}_{\tilde{y}}}^\dagger.$$

This leaves $M$ overparameterized and we can choose to give it the structure
$$M = \begin{bmatrix} M_{x|m} + M_m & -M_m \\ -M_m & M_m \end{bmatrix}$$

with
$$M_{x|m} = \beta^{-1} Z$$
$$M_m = \beta^{-1}(C_{y;x}^\intercal K_{x;y|m}^\intercal Z K_{x;y|m} C_{y;x} - Z).$$